\documentclass[10pt, aps,prc,twocolumn,superscriptaddress,preprintnumbers,
amsmath, 
floatfix,
longbibliography,
nofootinbib
]{revtex4-1}
\usepackage[T1]{fontenc}
\usepackage[utf8x]{inputenc} 
\usepackage{adjustbox}          % Used to adjust figure frames
\usepackage[caption=false]{subfig}
\usepackage{url}
\usepackage{color}
\usepackage{float}
\usepackage[pdftex,colorlinks=true, linkcolor = blue, citecolor=blue,urlcolor=blue, bookmarksnumbered=true, bookmarksopen=true]{hyperref}
\usepackage{longtable}
\usepackage{amsfonts}
\usepackage{dsfont}
\usepackage{wrapfig,bm} 
\usepackage[normalem]{ulem}
\usepackage{MnSymbol}
\usepackage{float}

\newcommand{\beq}{\begin{equation}}
\newcommand{\eeq}{\end{equation}}
\newcommand{\bea}{\begin{eqnarray}}
\newcommand{\eea}{\end{eqnarray}}

\begin{document}
\title{Entanglement entropy, single-particle occupation probabilities, and short-range correlations}
  
\author{Aurel Bulgac}% 
%\email{bulgac@uw.edu}%
\affiliation{Department of Physics,%
  University of Washington, Seattle, Washington 98195--1560, USA}
   
\date{\today}

\begin{abstract}
For quantum many-body systems with short-range correlations (SRCs), the intimate relationship between their magnitude, the behavior 
of the single-particle occupation probabilities at momenta larger than the Fermi momentum, and the entanglement entropy is a new qualitative 
aspect not studied and exploited yet.   A large body of recent condensed matter studies indicate that the time evolution of the entanglement entropy describes the 
non-equilibrium dynamics of isolated and strongly interacting many-body systems, in a manner similar to the Boltzmann entropy, 
which is strictly defined for dilute and weakly interacting many-body systems.
Both theoretical and experimental studies in nuclei and cold atomic gases 
have shown that the fermion momentum distribution has a generic behavior   
$n(k)=C/k^4$ at momenta larger than the Fermi momentum, due to the presence of SRCs, with approximately 
20\% of the particles having momenta larger than the Fermi momentum. The presence of the long momentum tails in the presence of
SRCs changes the textbook relation between the single-particle kinetic energy and occupation probabilities, $n_\text{mf}(k)    = {1}/\{ 1+\exp\beta[\epsilon(k)-\mu]\}$
for momenta very different form the Fermi momentum, particularly for dynamics processes.  
SRCs induced high-momentum tails of the single-particle occupation probabilities
increase the entanglement entropy of fermionic systems, which in its turn affects the dynamics of many nuclear reactions, such as 
heavy-ion collisions and nuclear fission.

\end{abstract}  

\preprint{NT@UW-22-04}

\maketitle   

Short-range correlations (SRCs) are defined as correlations between the constituents of a quantum many-body system at interparticle separations smaller than
the average separation between its constituents, which is of order $1/\sqrt[3]{n({\bm r})}$, where $n({\bm r})$ is the number density. Such interparticle separations correspond  to particle momenta larger than  the local Fermi momentum $\hbar k_F({\bm r})= \hbar \sqrt[3]{3\pi^2 n({\bm r})}$, when the particle motion is essentially unperturbed by the mean field. A great example was discussed by \textcite{Landau:1947}, when he 
discussed the dispersion relation between the (quasi)particle kinetic energy and its momentum 
$\varepsilon ({\bm p})$ in superfluid helium II, where he identified three branches of this spectrum: the phonon branch, the roton branch, and at higher momenta essentially the atom free motion. If the SRCs are strong, their role should appear at relatively small momenta, 
close to the Fermi momentum $\hbar k_F$
and the fraction of the quasiparticles with 
such momenta can be  significant. In particular,   the single particle 
occupation probability for Landau quasiparticles is given by the textbook formula~\cite{Abrikosov:1963},
apart from the trivial energy shift due the presence of the mean field, 
\begin{align}
& n_\text{mf}(k)    = \frac{1}{ 1+\exp\beta[\epsilon(k)-\mu] }, \label{eq:mf}
\end{align} 
where $\beta =1/T$ is the inverse temperature, $\mu$ is the chemical potential, and $\epsilon(k)={\hbar^2k^2}/{2m}$ 
is the kinetic energy.   
This formula is valid only in a relatively small energy interval around the Fermi level, strictly speaking only for systems in (quasi)equilibrium 
and it is in agreement with kinetic theory in the long time limit and quasi-locally for dilute and weekly interacting systems. Its range of 
validity and accuracy are however not established. Decades long studies of fermion 
momentum distribution in nuclei and cold atomic gases, where SRCs are important, 
require a more sophisticated approach. The relation between the SRCs and the entanglement entropy were discussed recently in
Refs.~\cite{Bulgac:2022,Bulgac:2022c,Bulgac:2022d,Pazy:2023}.

While it appears that only recently it was realized that quantum entanglement and superposition are equivalent concepts~\cite{Aubrun:2022}, 
with a little bit of effort one can easily convince oneself that the century old two-slit experiment due to Thomas Young,  
where superposition and coherence (thus entanglement) was crucial, the EPR paradox~\cite{Einstein:1935}, the 
entanglement introduced by Schr{\" o}dinger discussed many times by other authors \cite{Schrodinger:1935,Bell:1964,Wineland:2013}, 
and short-rage correlations(SRCs) in quantum systems where they are present, are intimately related phenomena. 
Two particles in a many-body system, interacting with forces with a range much shorter than the average particle separation, become naturally
entangled as in the situation discussed by \textcite{Einstein:1935}, 
when two particles are practically isolated from the rest of the Universe and retain the memory of the 
moment of ``creation'' of their initial state, a long time after they are fully spatially separated.  
Many-body systems with SRCs accordingly become entangled 
at all energies, irrespectively of whether the system is in equilibrium or not, and the measure of entanglement needs to be quantified.   
The system entanglement entropy directly affects 
the particle momentum distribution and its non-equilibrium dynamics~\cite{Bulgac:2022,Bulgac:2022c}. 
Isolated quantum systems in a pure state for which either von Neumann or Shannon entropy vanishes, 
but not necessarily in an equilibrium state, will evolve and its entanglement entropy will naturally in the long run describe their equilibration~\cite{Calabrese:2005,Calabrese:2006,Alba:2017,Milburn:1997,Vidal:2003,Korepin:2004,Kitaev:2006,Levin:2006,Li:2008,
Chuchem:2010,Pal:2010,Bardarson:2012,Cohen:2016,Abanin:2019,Sinha:2020,Wimberger:2021, 
Del-Maestro:2021,Del-Maestro:2022, Liu:2022,Mueller:2022,Schneider:2022}, 
similarly to the Boltzmann entropy for weakly interacting and dilute systems as discussed by \textcite{Boltzmann:1872,Nordheim:1928, Uehling:1933}.
Entanglement thus becomes a measure of both mean field  and short-range correlations as well~\cite{Bulgac:2022c}.

The single-particle momentum distribution can be extracted from the one-body density matrix
\begin{align}
n(\xi |\zeta) =\langle \Phi |\psi^\dagger(\zeta)\psi(\xi) |\Phi\rangle,    \label{eq:dens}
\end{align}
where $|\Phi\rangle$ is in general a time-dependent many-body wave function and
$\xi=({\bm r},\sigma, \tau),\, \zeta= ({\bm r}',\sigma',\tau')$ stand for the spatial, spin, and isospin coordinates.
Since the emphasis will be on the spatial properties, the spin and isospin degrees of freedom will be suppressed in ensuing equations. In 
many-body systems the density matrix typically is characterized by different spatial scales in the coordinates
${\bm R}= ({\bm r}+{\bm r}')/2$ and ${\bm s} = {\bm r}-{\bm r}$. 
The momentum distribution, obviously related to the Wigener distribution~\cite{Wigner:1932},  is defined~\cite{Benhar:1993} 
using Eq.~\eqref{eq:dens} for any many-body wave function
\begin{align}
&n({\bm k}) = \sum_{\sigma,\tau}\int d^3rd^3r'\, n({\bm r},\sigma,\tau|{\bm r'},\sigma,\tau) e^{-i{\bm k}\cdot ({\bm r}-{\bm r'})},\label{eq:nk0}
\end{align}
where $\int \tfrac{d^3k}{(2\pi)^3} n({\bm k})=A$ and $A=N+Z$ is the atomic number.
The properties of the nucleon momentum distribution have been investigated 
for decades~\cite{Levinger:1951,Cavedon:1982,Sartor:1980,Frankfurt:1981,
Frankfurt:1988,Benhar:1993,Pandharipande:1997,Quint:1986,Quint:1987,Ciofi-degli-Atti:1996,
Piasetzky:2006,Sargsian:2005,Schiavilla:2007,Bisconti:2007,Arrington:2012,Sargsian:2014,Hen:2014,Rios:2014,Hen:2017,
Zwerger:2011,Braaten:2011,Stewart:2010,Castin:2011,Anderson:2015,Porter:2017,Tan:2008a,
Tan:2008b,Tan:2008c,Yang:2019,Aumann:2021,Tropiano:2021,Tropiano:2022,Arrington:2022}. \textcite{Sartor:1980} have shown in 1980 that the momentum distribution of a dilute Fermi system 
is characterized by the presence of very long momentum tails $n(k)\propto 1/k^4$ at large momenta. 
Shina Tan~\cite{Tan:2008a,Tan:2008b,Tan:2008c} later proved that in the momentum interval $1/|a| \ll  k \ll 1/r_0$, where $a$ and $r_0$ are 
the $s$-wave scattering length and effective range respectively, the momentum distribution has the behavior $n(k)\approx C/k^4$. George Bertsch 
pointed in 1999 that dilute neutron matter~\cite{Baker:1999,Zwerger:2011} is exactly such a system.
Subsequent both theoretical and experimental studies for nuclear systems~\cite{Frankfurt:1981,Frankfurt:1988, Ciofi-degli-Atti:1996,
Piasetzky:2006,Sargsian:2005,Schiavilla:2007,Sargsian:2014,Hen:2014,Rios:2014,Hen:2017} 
 and for cold fermionic atom systems~\cite{Stewart:2010,Zwerger:2011,Braaten:2011,Castin:2011,Anderson:2015,Doggen:2015,Porter:2017} 
confirmed these predictions, even in cases where the interaction has quite a complex character, as in the case of a nuclear tensor interaction. 
The important conclusion of these studies was that approximately 20\% of 
the spectral strength is found for momenta $k>k_F$. As was mentioned by many ``{\it A crucial feature of the Tan relations is the fact that
they apply to any state of the system, e.g., both to a (normal) Fermi-liquid or to a superfluid
state, at zero or at finite temperature and also in a few-body situation}.''~\cite{Tan:2008a,Tan:2008b,Tan:2008c,Zwerger:2011,Castin:2011,Braaten:2011}.
 While nuclear studies were performed for understandable reasons only for the ground states 
of the systems, the experimental and theoretical results for cold fermionic atoms were obtained both at zero and finite temperatures, 
confirming Shina Tan's~\cite{Tan:2008a,Tan:2008b,Tan:2008c} prediction that the $n(k)=C/k^4$ behavior is in fact
generic for strongly interacting many-fermion systems, and thus a feature of such systems in both equilibrium and out of equilibrium.
In Refs.~\cite{Bulgac:2022c,Bulgac:2022} this aspect is illustrated 
for the case of highly excited fission fragments, with temperatures well above the critical temperature, where superfluid correlations are absent.   

\begin{figure} [h]
\includegraphics[width=0.9\columnwidth]{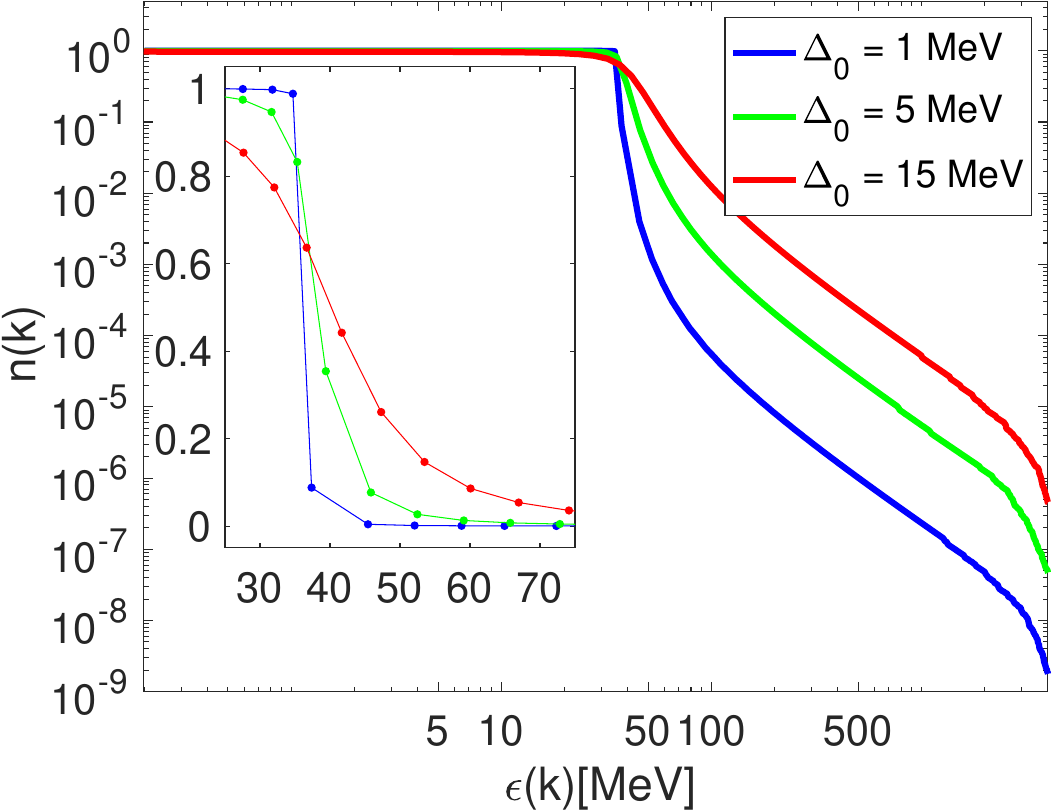}
\caption{ \label{fig:occ}  
Typical behavior of the  $s$-wave canonical occupation probabilities in the case of pairing in a finite nuclear system. 
Three different amplitudes of the pairing field were considered here $\Delta_0 = $ 1, 5, and 15 MeV~\cite{Bulgac:2022c}, 
the last one comparable to the strength of the pairing field in the unitary limit~\cite{Zwerger:2011}.
For $\epsilon(k)> 75$ MeV
the momentum occupation probabilities have the behavior $n(k)=C/k^4\propto 1/[\epsilon(k)]^2$. The UV-cutoff  
is determined by $\epsilon(k)$ of the highest canonical state with a wave function located inside the system. 
The states with $\epsilon(k)$ beyond the UV-cutoff are not expected to be physically relevant~\cite{Bulgac:2022c}, 
as their corresponding canonical occupation probabilities vanish in the continuum limit. 
In the inset, the occupation probabilities $n(k)$, shown in a linear scale,
have an expected Bardeen-Cooper-Schrieffer~\cite{Bardeen:1957}.}
\end{figure}

Typically one discusses the angle averaged momentum distribution  $n(k)=\int d\Omega_k n({\bm k})$, which can be 
evaluated by constructing the eigenvalues and eigenfunctions
 \begin{align}
\!\!\!\sumint_\zeta \,n(\xi|\zeta) \phi_\alpha (\zeta)\!=\!n_\alpha \phi_\alpha(\xi), 
\, n(\xi |\zeta) \!=\!\sum_\alpha\phi_\alpha (\xi)n_\alpha \phi_\alpha^* (\zeta) \label{eq:can}
\end{align}
known as the canonical basis in the case of the mean field Hartree-Fock-Bogoliubov approximation~\cite{Ring:2004}
or natural orbitals~\cite{Lowdin:1956,Lowdin:1956a} in general and evaluating
\begin{align} 
\epsilon(k) =  \left \langle  \phi_\alpha  \left | -\frac{\hbar^2\Delta }{2 m} \right | \phi_\alpha \right  \rangle = \frac{\hbar^2{\bm k}^2}{2m} \label{eq:ekin}
\end{align} 
and thus relating the occupation probability $n_\alpha=n(k)=n(\epsilon(k))$ with $\epsilon(k)$. (N.B. Obviously, the spectrum of Eq.~\eqref{eq:can} 
does not depend  on the specific representation, either coordinate, momentum, etc.)  
In saturating system, such as nuclei, the magnitude of the wave vector ${\bm k}$ 
is relatively well defined, up to corrections arising from surface effects~\cite{Brack:1997},
and the semiclassical quantization approach reproduces with very good accuracy 
single-particle energies and shell structure, an approximation going back to Bohr's model of the hydrogen atom.
For superfluid systems, if the pairing potential is local, then there is always a range of the wave vectors $k$ in 
which $n(k)=C/k^4$~\cite{Bulgac:2022c}, see Fig.~\ref{fig:occ}.
The presence of the high momentum tails $n(k)=C/k^4$ is clearly incompatible with Eq.~\eqref{eq:mf}, 
which decays exponentially when $k\rightarrow \infty$. Note that 
\begin{align} 
&n({\bm k})=\sum_{\alpha,\sigma,\tau} n_\alpha |\phi_\alpha({\bm k},\sigma,\tau)|^2, \\ 
&\phi_\alpha({\bm k},\sigma,\tau) = \int d^3r e^{-i{\bm k}\cdot{\bm r}}\phi_\alpha({\bm r},\sigma,\tau).
\end{align} 

%%%%%%%%%%%%%%%%%%%%%%%%%%%%%%%%%%%%%%%%%%%%

The SRCs, which modify in a qualitative manner the behavior the dependence of the occupation probabilities as a 
function of their kinetic energy, lead to significant changes of the entropy. This aspect can be appreciated in a much more ``down-to-earth''
language. In practice, when increasing the spatial resolution, and thus opening new channels and allowing higher momenta 
to actively participate in the dynamics, the system will always take advantage of new ``open roads'' and 
the the wave functions will spread naturally over a larger part of the Hilbert space. The entropy
is simply a measure of the available and 
allowed states into which the system can dynamically evolve. As mentioned in introduction, for isolated quantum many-body system 
the entanglement entropy describes the non-equilibrium dynamics~\cite{Calabrese:2005,Calabrese:2006,Alba:2017,Milburn:1997,Vidal:2003,Korepin:2004,Kitaev:2006,Levin:2006,Li:2008,
Chuchem:2010,Pal:2010,Bardarson:2012,Cohen:2016,Abanin:2019,Sinha:2020,Wimberger:2021, 
Del-Maestro:2021,Del-Maestro:2022, Liu:2022,Mueller:2022,Schneider:2022}, similarly to the Boltzmann entropy~\cite{Boltzmann:1872,Nordheim:1928,Uehling:1933}.
As recently discussed, the entanglement entropy is also a natural measure of the complexity of the wave-function of a quantum many-body system~\cite{Bulgac:2022c}.
The complexity of a many-body wave function $|\Phi\rangle$~\cite{Bulgac:2022c} can be quantified by evaluating the 
orbital entanglement or quantum Boltzmann one-body 
entropy~\cite{Nordheim:1928,Uehling:1933,Bulgac:2022,Klich:2006,Amico:2008,
Horodecki:2009,Haque:2009,Eisert:2010,Boguslawski:2014,Robin:2021}
\begin{align} 
S = - g\sum_\alpha \left [n_\alpha\ln n_\alpha  + (1-n_\alpha )\ln(1-n_\alpha ) \right \}, \label{eq:ent}
\end{align}
where the Boltzmann constant is $k_B=1$ when the temperature is measured in energy units, 
and $g$ is spin-isospin degeneracy factor. The set of occupation numbers $\{n_\alpha, 1-n_\alpha\}$ are known as entanglement spectrum and obviously
cary more information than the entanglement entropy~\cite{Li:2008}.
A very transparent derivation of the orbital entanglement entropy for fermion systems was given in Ref.~\cite{Gigena:2015} 
using the decomposition, similar to Schmidt decomposition or tensor of the many-body wave function discussed a long time by \textcite{Coleman:1963}
\begin{align} 
\!\!\!|\Phi\rangle = a^\dagger_\alpha a_\alpha |\Phi\rangle + a_\alpha a^\dagger_\alpha |\Phi\rangle = 
\sqrt{n_\alpha} |\Phi_\alpha \rangle +\sqrt{1-n_\alpha}|\Phi_{\overline{\alpha}}\rangle,
\end{align} 
where $a^\dagger_\alpha, a_\alpha$ are creation and annihilation fermionic operators. 
This entropy vanishes 
exactly for an isolated system in a pure Hartree-Fock wave function, thus non-interacting particles, and it is different from zero 
only in the presence of residual interactions, both for ground and excited states. Cold fermionic gases where the only interaction is a zero-range 
is perhaps the  most simple example of a strongly interacting system where for any scattering length any exact many-body state 
has SRCs~\cite{Tan:2008a,Tan:2008b,Tan:2008c} and the entanglement entropy $S$~\eqref{eq:ent} is non-vanishing for any state, 
irrespective of whether a pairing condensates is present or not.
In the limit of a dilute system with weak particle interactions the entanglement entropy $S$ is a very good approximation of the 
thermodynamic entropy in both classical~\cite{Boltzmann:1872} and quantum semiclassical limits~\cite{Nordheim:1928,Uehling:1933,Abrikosov:1963}.
This entanglement entropy $S$ is a measure of the amount of the many-body correlations beyond a pure 
Slater determinant~\cite{Bulgac:2022c}, for which $S\equiv 0$. 
The orbital entanglement entropy $S$ defined in Eq.~\eqref{eq:ent} is related (though it is not identical) to the 
Shannon entropy~\cite{Bengtsson:2017} familiar in quantum information science, see a discussion in Ref.~\cite{Bulgac:2022c}.
The entanglement entropy never vanishes~\cite{Srednicki:1993,Bengtsson:2017} for an interacting system in a pure state, and in particular in its ground state.
The set of  $-\ln n_k$ is also known as the entanglement spectrum~\cite{Li:2008}.  

\begin{figure} 
\includegraphics[width=0.9\columnwidth]{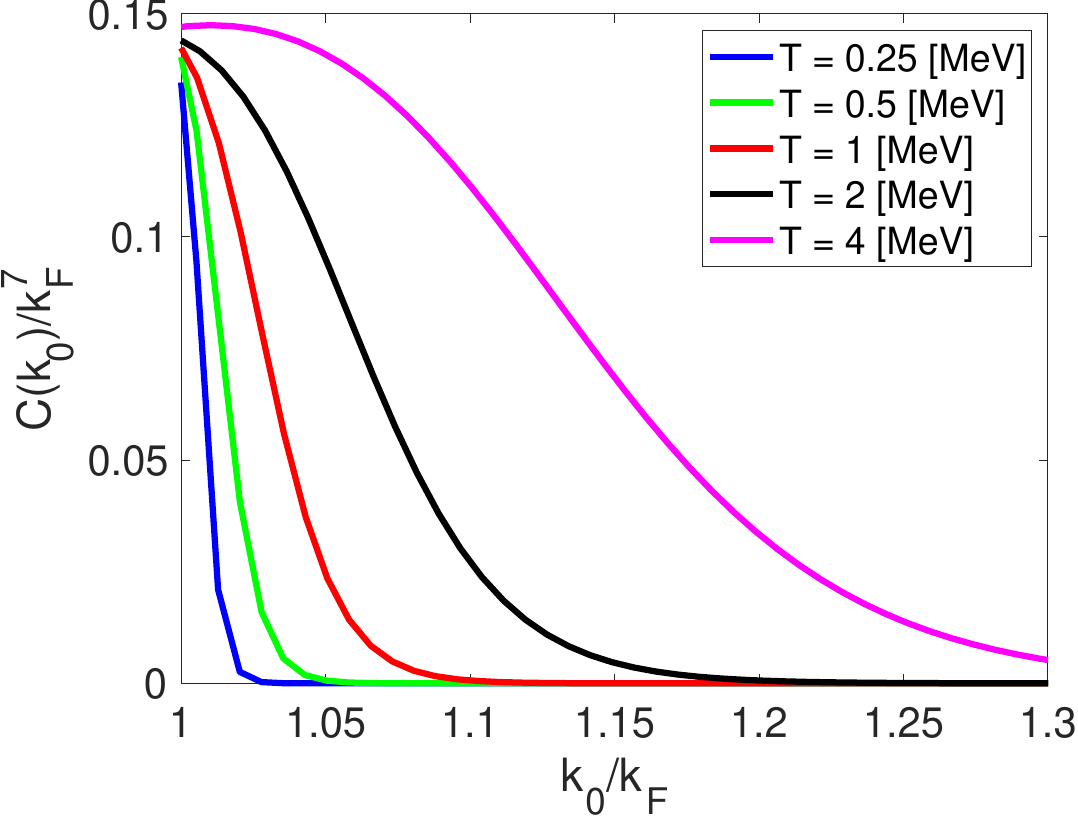}
\caption{ \label{fig:C}  
The dependence of the dimensionless ``contact'' $C(k_0)/k_F^7$ 
on the choice of momentum scale $k_0$ extracted by imposing the 
normalization condition  Eq.~\eqref{eq:norm} on the occupation probability $n(k)$,   
for a thermal mean field distribution, see Eq.~\eqref{eq:nk} and where the temperature $\beta =1/T$,
increases from the lowest to the highest curve. }
\end{figure}

\begin{figure} 
\includegraphics[width=0.9\columnwidth]{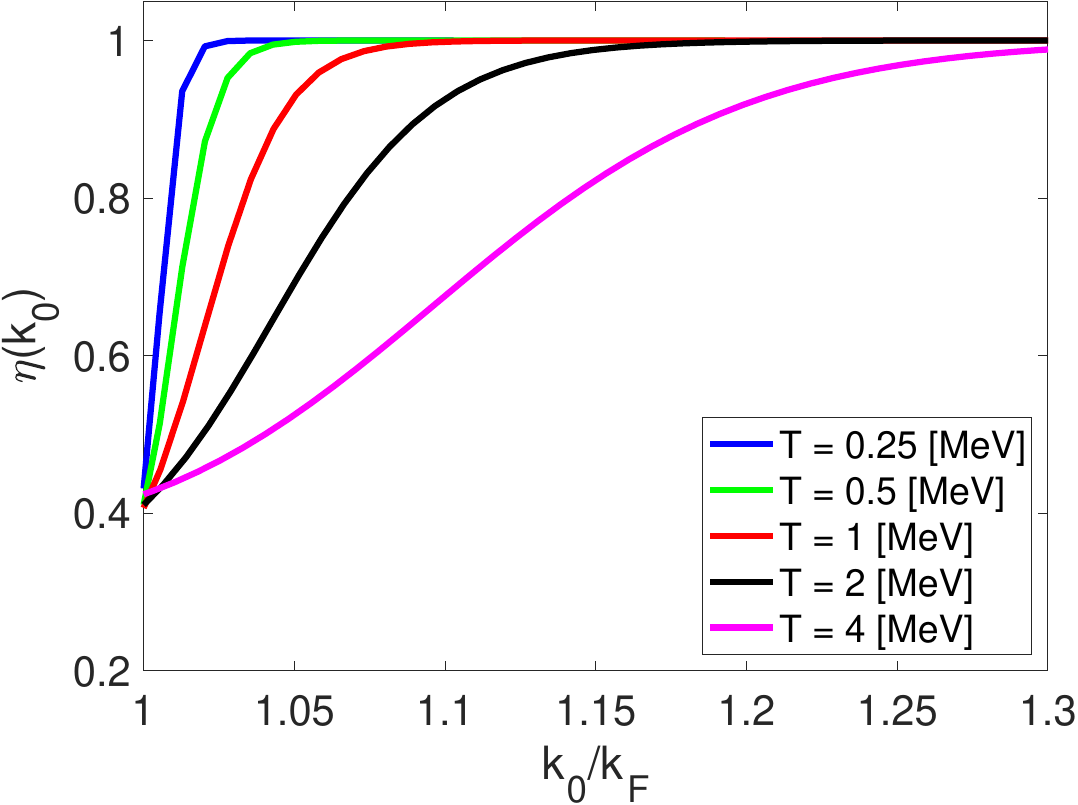}
\caption{ \label{fig:eta}  
The dependence of $\eta(k_0)$ on the choice of momentum scale $k_0$ extracted from Eq.~\eqref{eq:eta},  for different temperatures
when $n_\text{mf}(k)$ is a thermal distribution for free fermion gas. For $k_0>1.05 k_f$ the temperature decreases from the lowest to the highest curve. }
\end{figure}

\begin{figure}[h]
\includegraphics[width=0.9\columnwidth]{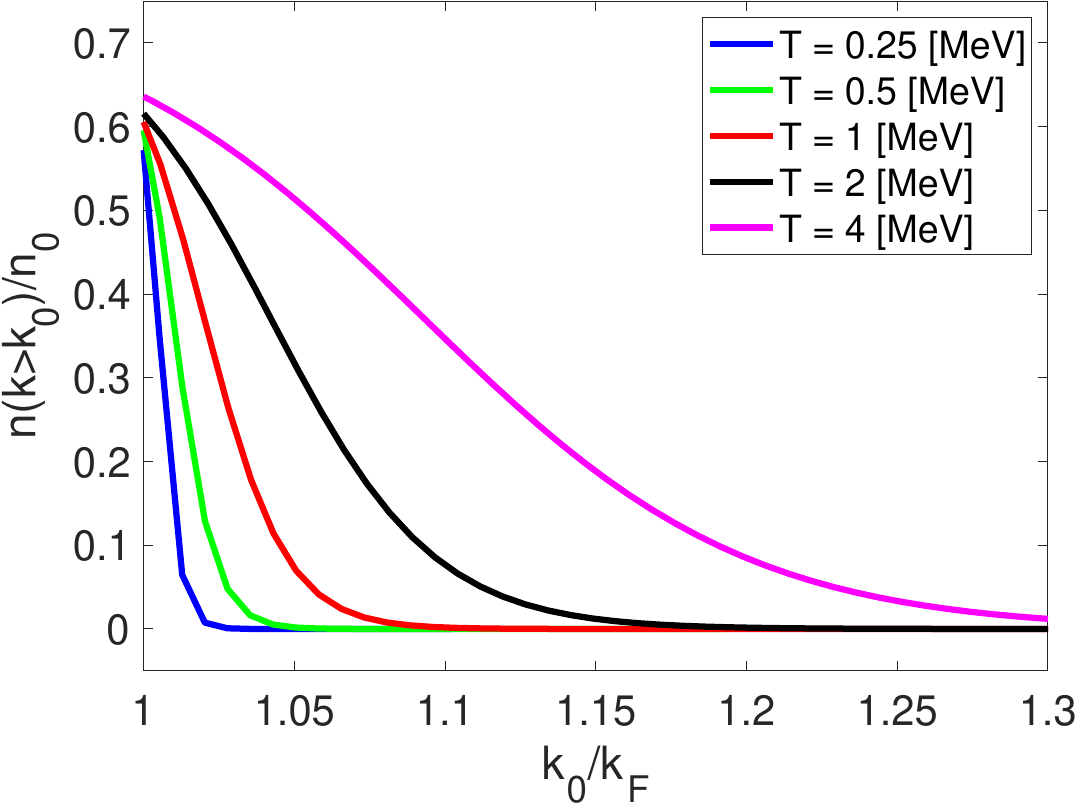}
\caption{ \label{fig:alpha}  
The fraction of the particles $n(k>k_0)/n_0 $ with momenta $k\ge k_0$, where $n(k) \propto 1/k^4$. 
The curves rank by temperature.}
\end{figure}

When accounting for the SRCs the nucleon momentum distribution can be parametrized with a simple model~\cite{Hen:2014}
\begin{align} 
\!\!\!\!\!\!\!\!\!\! &n(k) = \eta(k_0) 
\left \{
\begin{matrix} 
n_\text{mf}(k),                                         \quad &\text{if}\quad &k\leq k_0\\
n_\text{mf}(k_0)k_0^4/{k^4}, \quad &\text{if}\quad &k_0<k<\Lambda
\end{matrix} 
\right. , \label{eq:nk} \\
&C(k_0)= \eta(k_0) n_\text{mf}(k_0)k_0^4,\\
&n_0 = g\int_{k<\Lambda} \frac{d^3k}{(2\pi)^3}n(k) =\frac{gk_F^3}{6\pi^2},\label{eq:norm}
 \end{align}
where $C(k_0)$  the ``contact'' introduced by Shina Tan~\cite{Tan:2008a,Tan:2008b,Tan:2008c}
and $n_0\approx 0.16$ fm$^{-3}$ is the saturation density of symmetric nuclear matter.
The SRCs part of the momentum distribution $n(k)$ is clearly a beyond mean field feature in nature. 
Since so far no argument has been suggested in literature that 
a discontinuity of $n(k)$ at $k_0$ might occur, it is reasonable to assume that $n(k)$ is continuous and thus there is a simple 
interpretation of the contact $C(k_0)=\eta(k_0)n_\text{mf}(k_0)k_0^4$ in terms of a single parameter $k_0$, see Fig.~\ref{fig:C}. 
The presence of a cusp at $k_0$ in the adopted parameterization of $n(k)$ introduces only relatively small numerical corrections.
%%%%%%%%%%%%%%%%%%%%%%%%%%%%%%%%%%%%%%%%%%%%
%%%%%%%%%%%%%%%%%%%%%%%%%%%%%%%%%%%%%%%%%%%%
The normalization constant $\eta(k_0)$, which characterizes the depletion of the Fermi sea due to residual interactions,
is obtained from the condition Eq.~\eqref{eq:norm}, see Fig.~\ref{fig:eta},
\begin{align}  \label{eq:eta}
\eta(k_0) = \frac{\int_{0}^\Lambda dk\, k^2 n_\text{mf}(k)}{ k_0^3n_\text{mf}(k_0) +\int_{0}^{k_0} dk\, k^2 n_\text{mf}(k)}.
\end{align}
Assuming that $\Lambda=\infty$ and
\begin{align}
 n_0 = g\int \frac{d^3k}{(2\pi)^3} n_\text{mf}(k).
\end{align}  
a lower limit for $\eta(k_0)$ can be obtained in the case of a free Fermi gas at zero temperature by choosing $k_0=k_F$.
One can now evaluate the fraction of the particles with momenta greater than $k_0$, see Fig.~\ref{fig:alpha},
\begin{align}
\frac {n(k>k_0)}{n_0}=\frac{3\eta n_\text{mf}(k_0)k_0^3}{k_F^3}=\frac{3C(k_0)}{k_0k_F^3},
\end{align}  
which can reach quite large values. 
It is not my goal here to select the best choice for the mean field momentum probability distribution, as that should be 
decided in accurate microscopic calculation, specific for various systems~\cite{Sargsian:2005,Piasetzky:2006,Schiavilla:2007,
Zwerger:2011,Braaten:2011,Castin:2011,Drut:2011,
Arrington:2012,Sargsian:2014,Anderson:2015,Porter:2017,Hen:2017,Richie-Halford:2020}.

%%%%%%%%%%%%%%%%%%%%%%%%%%%%%%%%%%%%%%%%%%%%

The UV-momentum cutoff  $\Lambda$ of the momentum distribution $n(k)$ is effective field theory in nature 
and is determined by the internal structure of the nucleons.
In the limit $k_0\rightarrow \infty $ the ``contact'' $C$  
naturally vanishes. The specific value of the 
``contact'' $C$ is defined by the temperature $T$, the specific system under consideration,  and the system specific momentum scale   
$k_0 $~\cite{Sartor:1980,Frankfurt:1981,Frankfurt:1988,Ciofi-degli-Atti:1996,Piasetzky:2006,Bulgac:2005,Tan:2008a,Tan:2008b,Tan:2008c,
Astrakharchik:2004,Chang:2004,Sargsian:2005,Schiavilla:2007,Hen:2014,Stewart:2010,Drut:2011,Arrington:2012,
Sargsian:2014,Anderson:2015,Porter:2017,Hen:2017,Richie-Halford:2020}.

 %%%%%%%%%%%%%%%%%%%%%%%%%%%%%%%%%%%%%%%%%%%%

The momentum distribution $n(k)$ has thus two components, 
the mean field and the SRCs components, which can be clearly identified experimentally~\cite{Stewart:2010,Hen:2014,Hen:2017}
by identifying the regime $k_0 < k<\Lambda$ where $n(k)\approx  C(k_0)/k^4$ is valid. Below $k<k_0$ the $n(k)$ can be then 
identified with a mean field contribution, up to the overall renormalization 
constant $\eta(k_0)\leq 1$. The constant $\eta(k_0)$ 
is uniquely determined by the normalization condition Eq.~\eqref{eq:norm} and the mean field 
component $n_\text{mf}(k)$, which is determined in a typical mean field or Density Functional Theory (DFT)~\cite{Hohenberg:1964,Kohn:1965}.
The momentum distribution $n(k)$ depends on a single parameter $k_0> k_F$,
that can be determined either experimentally or from DFT with a sufficiently large UV-momentum cutoff or from 
another accurate many-body calculation, when pairing correlations are also taken 
into account~\cite{Bulgac:2022,Bulgac:2022a}. 
 
\begin{figure}[h] 
\includegraphics[width=1.0\columnwidth]{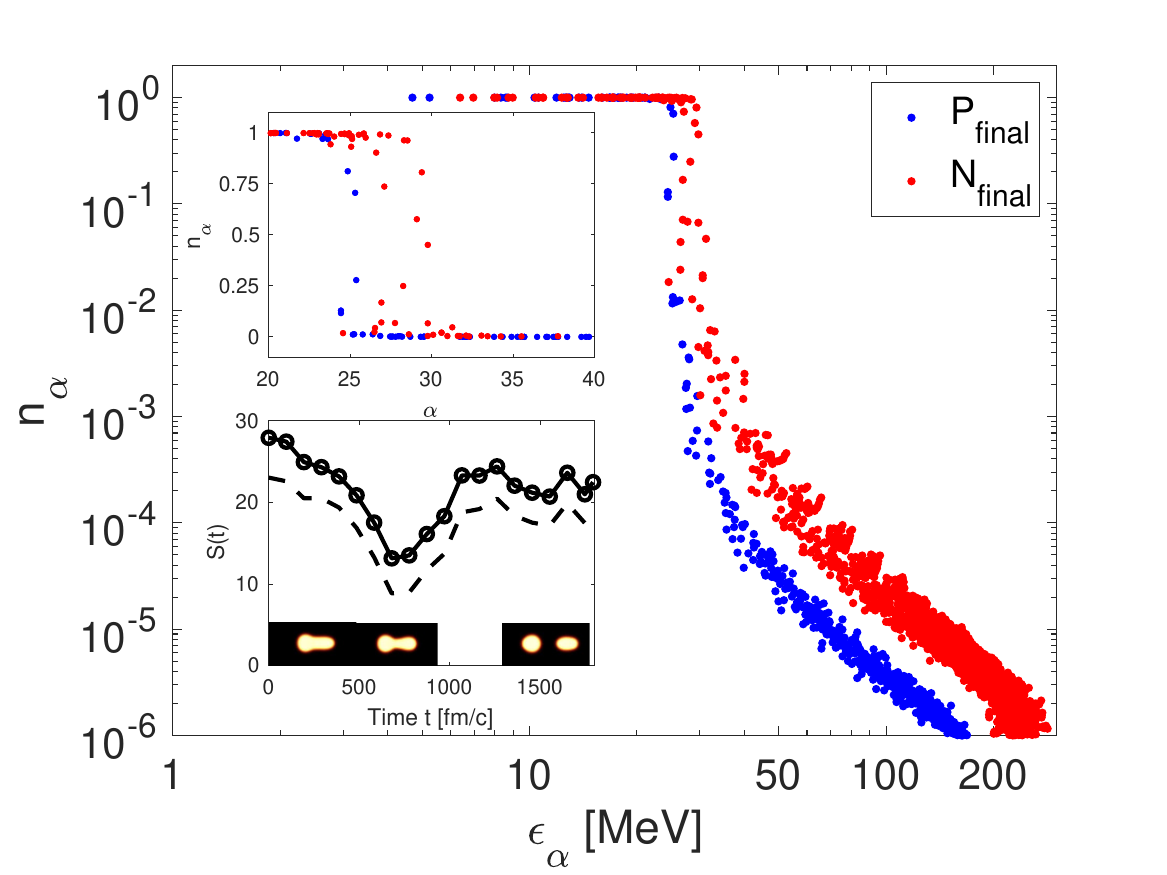}
\caption{ \label{fig:n_ek}   
The final proton and neutron canonical occupation probabilities 
$n_\alpha= n(k)$ extracted from the TDDFT extended to superfluid systems 
treatment of  induced fission reaction $^{235}$U(n,f) as a function of $\epsilon_\alpha=\epsilon(k)$. The upper inset shows a 
small  energy interval near the Fermi level. Above $\epsilon_\alpha \approx 50$ MeV one can see a 
clear power law behavior compatible with theory prediction $n(k)\propto 1/[\epsilon(k)]^2$.
The initial state was the compound nucleus close to the top of the outer fission barrier at $t=0$ fm/c 
and in the final state the fission fragments are spatially separated by 
$\approx 30$ fm  at $t=1,700$ fm/c~\cite{Bulgac:2022c}.  The non-equilibrium 
time-evolution of the orbital entanglement entropy $S(t)$ is shown in the lower inset, with solid and dashed lines corresponding 
to no nucleon number projections and with nucleon number projections respectively~\cite{Bulgac:2022c}. }
\end{figure}

In the classic monograph~\cite{Abrikosov:1963} there is a somewhat hard to interpret sentence, stating that the  
dependence of the occupation probabilities on the quasiparticle energies $\epsilon$  is {\it a very complicated implicit 
definition of $n(\epsilon)$} (see Eq.~(2.6) in Ref.~\cite{Abrikosov:1963} and the 
corresponding explanations), whereas $n(\epsilon)$ is clearly a well defined function of $\epsilon$, see Eq.~\eqref{eq:mf}. 
In Fig.~\ref{fig:n_ek} I show the dependence 
of the canonical occupation probabilities  $n_\alpha=n(\epsilon(k))$  in case of induced fission $^{235}$U(n,f), 
extracted from the time-dependent DFT (TDDFT) approach extended
to superfluid systems and applied to this non-equilibrium process~\cite{Bulgac:2013a,Bulgac:2019,Bulgac:2016,Bulgac:2019c,Bulgac:2020}. 
These results show the momentum distribution of two hot emerging fission fragments, with a separation in space $\approx 30$ fm, and 
at  temperatures larger than the critical temperature $T_c \approx 0.5$ MeV
for which the pairing gaps vanish, and which demonstrate the 
clear presence of proton-proton and neutron-neutron SRCs~\cite{Bulgac:2022}. 
In the dynamics of isolated systems the time evolution of the entanglement entropy plays the role of thermodynamic entropy 
for local observables~\cite{Calabrese:2005,Calabrese:2006,Alba:2017}, is shown in the lower inset of Fig.~\ref{fig:n_ek}. Note that at the initial time
the nucleus is at zero temperature, but the entanglement entropy does not vanish. For more details see Ref.~\cite{Bulgac:2022c}.
In these calculations nucleon momenta up to $p_\text{cut} =\hbar \pi/dx \approx 600 $ 
MeV/c (where the spatial resolution is $dx = 1$ fm) are present.
According to the prevalent interpretation of time-dependent mean field treatment of many fermion systems, 
only long-range correlations should be present, which obviously 
is not the case in TDDFT extended 
to superfluid systems~\cite{Bulgac:2013a,Bulgac:2019,Bulgac:2022,Bulgac:2022c}.
This dependence of $n(k)$ on $\epsilon(k)$, where the long-momentum tails are present,
is indeed a complicated implicit definition of the canonical occupation probabilities. The canonical basis set 
is the unique (gauge invariant) and at the same the minimal set of single-particle states to represent a many-body wave 
function~\cite{Lowdin:1956,Lowdin:1956a,Coleman:1963,Davidson:1972,Bulgac:2022c}. 
This  clarifies perhaps for the first time the meaning of the sentence quoted above and an equivalent of 
which I could not find in literature. The presence of the  infrared (IR) knee  at $\epsilon_\alpha\approx 40$ MeV is 
unequivocally a qualitative new feature, absent from the textbook definition~\cite{Abrikosov:1963} of a quasi-equilibrium distribution $n(\epsilon)$. 
%%%%%%%%%%%%%%%%%%%%%%%%%%%%%%%%%%%%%%%%%%%%

\begin{table}[ht]
\begin{tabular}{|l| r| r| r| r| r|} 
\hline \hline
             &$\tfrac{k_0}{k_F}$=1.3& $\tfrac{k_0}{k_F}$=1.2&  $\tfrac{k_0}{k_F}$=1.1&    $\tfrac{k_0}{k_F}$=1.05&         $\tfrac{k_0}{k_F}$= 1\\
\hline
T = 0.25 MeV&                 1&        1&        1&         1&  124.6\\ 
T = 0.5  MeV&                 1&        1&        1&    2.18&    60\\
T = 1     MeV&                 1&        1&   1.28&    8.92&    29.8\\
T = 2    MeV &                 1&    1.11&   4.01& 10.48&    14.9\\
T = 4     MeV&            1.27&    2.41&  5.57&   6.81&       7.60\\
\hline\hline         
 \end{tabular}
 \caption{ \label{tab:I} 
 The values of the ratio of the entropy of the system, in the presence of SRCs, evaluated with Eq.~\eqref{eq:nk}, over the entropy evaluated 
 in pure mean field approximation evaluated evaluated with Eq.~\eqref{eq:mf},
 for different values of the momentum scale $k_0$, is shown in columns 2-6.  }
 \end{table}

After a cursory analysis of 
Eq.~\eqref{eq:ent}, one is lead to the conclusion that due to the presence of the SRCs contribution  
this entanglement entropy likely exceeds in value the corresponding entanglement mean field entropy, see Table~\ref{tab:I}. 
The SRCs contribution to $n(k)$ has a very long power law tail, 
which would lead to $\eta(k_0)<1$, see Fig.~\ref{fig:eta}, and thus to an expected depletion of the occupation probabilities of 
the low-momentum states $k<k_0$, even at very low temperatures. This occupation probability depletion of the 
states with $k<k_0$ alone would lead to an increase of the corresponding contribution of these states to the entropy 
density of the system. At the same time, the long tails of the momentum distribution for $k>k_0$ would lead to a 
further increase of the entropy density, when compared to the mean field value.  Since $\Lambda \gg k_0$, the effect 
of considering the internal nucleon structure have likely a relatively small effect on the entropy, 
which is well converged when $n(\Lambda)\approx 10^{-7}$. Upon performing a projection on exact proton and neutron numbers
the many-body wave function is an exponentially large sum of Slater determinants 
(a typical shell-model or configuration interaction many-body wave function), 
the canonical/natural orbital occupation probabilities  remain largely unchanged~\cite{Bulgac:2022c,Bulgac:2022d} and thus the particle projected 
many-body wave function retains a very high degree of entanglement.
This many-body wave function is solution of the quantum equivalent  of the semiclassical Boltzmann equation~\cite{Bulgac:2022c}.

Pairing correlations alone lead to $1/k^4$ tails in the 
momentum distribution at all temperatures~\cite{Bulgac:2022,Bulgac:2022a}. Moreover, the dynamical pairing effects,
namely the presence of a pairing field, but the absence of a true pairing condensate at temperatures higher 
than the critical temperature, lead to the occupation of high-momentum states with $k>k_0$~\cite{Bulgac:2019c,Bulgac:2020,Bulgac:2022},  
even in time-dependent processes and for intrinsic excitation energies of nuclei corresponding to temperatures above the critical temperature.  
Since these pairing correlations 
in current nuclear simulations take into account only the $nn$ and $pp$ correlations~\cite{Bulgac:2022},
the effects of $np$ SRCs can be included in dynamical calculations by an extension of time-dependent DFT described in Ref.~\cite{Bulgac:2022},
are expected to be significantly larger.
Since entanglement entropy and many-body level density control the dynamics of an isolated quantum 
system~\cite{Calabrese:2005,Calabrese:2006,Alba:2017}, the level density in the presence
of SRCs exceeds the level density of the system in a simpler mean 
field approximation. The possibility that the momentum distribution may be time-dependent as well was not explicitly discussed here, 
only indirectly illustrated in the lower inset in Fig.~\ref{fig:n_ek},
it was definitely observed in time-dependent microscopic quantum studies~\cite{Bulgac:2022,Bulgac:2022c,Bulgac:2022d}.
The highly non-equilibrium nuclear fission $^{235}$U(n,f) illustrated here and in Refs.~\cite{Bulgac:2022c,Bulgac:2022d} in  arguably  
the largest many-body system studied so 
far~\cite{Milburn:1997,Vidal:2003,Korepin:2004,Kitaev:2006,Levin:2006,Li:2008,
Chuchem:2010,Pal:2010,Bardarson:2012,Cohen:2016,Abanin:2019,Sinha:2020,Wimberger:2021, 
Del-Maestro:2021,Del-Maestro:2022, Liu:2022,Mueller:2022,Schneider:2022},  
with aspects related to the widely studied topics of Hilbert space and many-body localization.
The presence of SRCs lead to qualitative changes of the entanglement properties, the complexity 
of the many-body wave functions, the single-particle occupation probabilities, and the dynamics of many-body
systems~\cite{Bulgac:2009,Bulgac:2011,Bulgac:2011a,Bulgac:2017,Bulgac:2019c,Bulgac:2020,
Magierski:2022,Bulgac:2022,Bulgac:2022c,Bulgac:2022d}.
Nuclear and cold atom systems present a unique opportunity to study time-dependent
non-equilibrium and entanglement properties of  strongly interacting fermions.  

\vspace{0.5cm}
        
{\bf Acknowledgements} \\

The funding 
from the US DOE, Office of Science, Grant No. DE-FG02-97ER41014 and
also the support provided in part by NNSA cooperative Agreement
DE-NA0003841 is greatly appreciated. 
This research used resources of the Oak Ridge
Leadership Computing Facility, which is a U.S. DOE Office of Science
User Facility supported under Contract No. DE-AC05-00OR22725. I thank 
M. Kafker and I. Abdurrahman for the help in computing the results presented in Figs.~\ref{fig:occ} and  \ref{fig:n_ek}. 

%%%%%%%%%%%%%%%%%%%%%%%%%%%%%%%%%%%%%%%%%%%%%

% These are needed to avoid a babel error.
\providecommand{\selectlanguage}[1]{}
\renewcommand{\selectlanguage}[1]{}

\bibliography{local_fission}

\end{document}